\documentclass[10pt,a4paper]{article}
\usepackage[utf8]{inputenc}
\usepackage[T1]{fontenc}
\usepackage[english]{babel}
\usepackage{newtxtext}
\usepackage{newtxmath}
\usepackage[fontsize=9.0bp]{scrextend}

\usepackage{amsmath}
\usepackage{amsfonts}
\usepackage{bm}
\usepackage{listings}

\usepackage{graphicx}
\usepackage[pdfa,pdfproducer={},pdfcreator={}]{hyperref}
\usepackage{url}

\usepackage{mathtools}
\mathtoolsset{showonlyrefs=true}

\usepackage{xcolor}
\usepackage{colortbl} 

\definecolor{bleudefrance}{rgb}{0., 0.37, 0.67}
\definecolor{colorEMG}{HTML}{c04559}
\definecolor{colorMone}{HTML}{009988}
\definecolor{colorMtwo}{HTML}{0077BB}
\definecolor{colorMthree}{HTML}{004488}

\hypersetup{
	colorlinks = true,
	citecolor ={bleudefrance},
	allcolors= {bleudefrance},
	linkbordercolor = {white},
}

\usepackage{tikz}

\newsavebox{\legendEMG}
\sbox{\legendEMG}{\raisebox{0.5ex}{\tikz{\draw[colorEMG, dashed, line width=0.9pt] (0,0) -- (0.4cm,0);}}}

\newsavebox{\legendMone}
\sbox{\legendMone}{\raisebox{0.5ex}{\tikz{\draw[colorMone, line width=0.9pt] (0,0) -- (0.3cm,0);}}}

\newsavebox{\legendMtwo}
\sbox{\legendMtwo}{\raisebox{0.5ex}{\tikz{\draw[colorMtwo, line width=0.9pt] (0,0) -- (0.3cm,0);}}}

\newsavebox{\legendMthree}
\sbox{\legendMthree}{\raisebox{0.5ex}{\tikz{\draw[colorMthree, line width=0.9pt] (0,0) -- (0.3cm,0);}}}

\usepackage[font=small,justification=RaggedRight]{caption}
\usepackage{microtype}
\microtypesetup{
	protrusion=true,
	expansion=true,
	tracking=false,
	spacing=true,       
	final
}
\usepackage{extdash}

\usepackage[small,compact]{titlesec}
\titleformat{\section}{\bfseries\sffamily\scshape\color{black}}{\arabic{section}}{1em}{\centering\MakeUppercase}
\titleformat{\subsection}{\raggedright\bfseries\sffamily\scshape\small}{\arabic{section}.\arabic{subsection}}{1em}{\MakeUppercase}
\titleformat{\subsubsection}{\centering\bfseries\sffamily\scshape\footnotesize}{\arabic{section}.\arabic{subsection}.\arabic{subsubsection}}{1em}{\MakeUppercase}

\titleformat{\section}
{\sffamily\bfseries\scshape\normalsize}
{\arabic{section}}
{1em}
{\raggedright\MakeUppercase} 

\titleformat{\subsection}
{\sffamily\bfseries\small} 
{\arabic{section}.\arabic{subsection}}
{1em}
{\raggedright}

\titleformat{\subsubsection}
{\sffamily\itshape\small}
{\arabic{section}.\arabic{subsection}.\arabic{subsubsection}}
{1em}
{\raggedright}

\titlespacing{\subsubsection}{0pt}{*4}{*1}
\titlespacing{\subsection}{0pt}{*4}{*1}
\titlespacing{\section}{0pt}{*4}{*1}

\usepackage{fancyhdr}
\usepackage[left=1.350cm, right=1.350cm, top=1.5cm, bottom=2.00cm]{geometry}
\usepackage[backend=bibtex,
isbn=true,
url=true,
eprint=false,
bibstyle=numeric,
sorting=none,
style=numeric-comp,
punctfont=false,
date=year,
maxbibnames=10,
minbibnames=10]{biblatex}

\DeclareFieldFormat{pages}{#1}

\AtEveryBibitem{%
	\clearfield{note}%
}
\AtEveryBibitem{%
	\clearfield{note}%
	\ifentrytype{article}{%
		\clearfield{url}%
		\clearfield{urldate}%
	}{}
	\ifentrytype{article}{%
		\clearfield{issn}
	}{}
}

\DeclareFieldFormat{doi}{%
	\textsc{doi}\addcolon\space
	\ifhyperref
	{\href{https://doi.org/#1}{#1}}
	{#1}}

\DeclareCiteCommand{\citenum}
{}
{\printfield{labelnumber}}
{}
{}
\bibliography{bibs}

\usepackage{ragged2e}
\usepackage{booktabs}
\usepackage{tabularx}

\usepackage{placeins}

\usepackage[misc]{ifsym}

\usepackage{float}
\restylefloat{table}

\usepackage{appendix} 

\usepackage{abstract}
\usepackage{orcidlink}

\usepackage{multicol} 

\begin{document}
		
	\title{\Large\textsf{\textbf{Chromatographic Peak Shape from a Stochastic-Diffusive Model with Multiple Retention Mechanisms: Analytic Time-Domain Expression and Derivatives						
	}}}
	
	\author{\small \textsf{{Hernán R. Sánchez}}$^{1}${\scriptsize\Letter}\orcidlink{0000-0003-1058-3396}}	
	\date{}
	
	\maketitle
	
	\begin{center}
		{\vspace{-0.5cm}\small 
			Instituto de Física de Líquidos y Sistemas Biológicos,UNLP-CONICET, La Plata, 1900, Argentina
			
			\footnotesize {\scriptsize\Letter} \textsf{\href{mailto:hernan.sanchez@quimica.unlp.edu.ar}{hernan.sanchez@quimica.unlp.edu.ar}}
		}
	\end{center}

	\vspace{-0.3cm}

\begin{abstract}
	A time-domain analytic expression for chromatographic peak shapes
	is derived within a stochastic-diffusive framework that incorporates
	axial diffusion (molecular and multipath/Eddy), finite initial
	spatial variance, a retention mechanism characterized by a high rate
	of short-duration events, and an arbitrary number of
	independent slow retention mechanisms, each characterized by its own
	rate of infrequent, long-duration events.
	A highly efficient evaluation scheme is derived for this expression.
	In the single-slow-mechanism case, it is two to four orders of
	magnitude faster than the previously available analytic route.
	Analytical derivatives with respect to all model parameters are also
	obtained, and each can be evaluated at computational cost comparable
	to that of the peak-shape expression.
	Illustrative fits to three literature peaks yielded full-profile
	RMSE values lower than those of the exponentially modified Gaussian
	in all tested cases, with minima ranging from 0.03 to 0.14 percent
	of peak height, compared with 0.43 to 5.57 percent for the reference
	model. Relative to the one-slow-mechanism formulation, allowing more
	than one slow mechanism produced a data-dependent improvement that
	exceeded one order of magnitude for one of the peaks.
\end{abstract}

	\vspace*{0.cm}
	
\begin{multicols}{2}
		\setlength{\parindent}{1.0em}

\section{Introduction}

A central objective in separation theory is to obtain mathematical expressions that represent chromatographic peak shapes. 
Such expressions enable the direct fitting of experimental peaks and, when derived from mechanistic models of intracolumn processes, the estimation of physically interpretable parameters.
Although phenomenological models parameterized by empirical shape descriptors offer the computational convenience required for routine analysis\cite{dimarco2001}, a longstanding limitation has been the lack of analytic time-domain expressions capable of reconciling practical tractability with mechanistic breadth.

Detailed process-level models are indispensable for simulation and design \cite{Bernau2022}, but their mathematical complexity precludes the derivation of such expressions.
Among mechanistic approaches, stochastic frameworks \cite{EyringGiddings1955,GiddingsCalvin1957,McQuarrie1963,Weiss1970,felinger1999stochastic} are particularly suited to this purpose, 
as they relate the single-analyte peak profile to the probability laws used to describe its transport through the mobile phase and its transient retention by the stationary phase. Under linear detection, the analyte-specific contribution to the chromatographic signal is proportional to the probability density function (PDF) of the single-particle elution time, which is approximated in theoretical treatments by the intracolumn residence-time distribution.

Until recently, however, the most comprehensive stochastic treatments of chromatographic retention were developed in the Laplace or Fourier domain, largely because the mathematical treatment of models involving multiple independent processes is considerably simpler in the transform domain~\cite{Dondi1986,CavazziniRemelliDondiFrancesco1997,Cavazzini1999,pasti2016}. Recovering the time-domain peak from such representations requires numerical inversion, which interposes an intermediate computational step between the transform-domain expression and the observable chromatographic peak.
This indirection obscures the correspondence between the mathematical expression and the peak profile. For peak fitting, this is a drawback because time-domain strategies such as fitting on a subset of the data points, or progressively refining the evaluation grid, do not transfer directly to transform-domain formulations.
Furthermore, relying on transform-domain representations has been shown to degrade the precision of parameter estimation under realistic chromatographic conditions\cite{felinger2011}.
These considerations help explain the limited practical use of such transform-domain stochastic formulations.

In a recent study by the author~\cite{Sanchez2026}, this gap was substantially narrowed by deriving a rapidly evaluable analytic time-domain expression for chromatographic peak shapes from a comprehensive stochastic–diffusive framework that unifies, within a single mathematical structure, all of the principal transport and retention processes involved in peak formation.
This combination of mechanistic breadth and practical computability makes the resulting expression well suited to direct peak fitting and to the estimation of transport and kinetic parameters with direct physical meaning.

The framework of Ref.~\cite{Sanchez2026} describes chromatographic peak formation in linear chromatography in terms of single-particle transport and retention processes. 
It incorporates advective–diffusive transport in the mobile phase, comprising molecular diffusion and multipath (Eddy) dispersion, together with the finite initial spatial variance of the injected band. Retention in the stationary phase is described through two independent mechanisms: a fast one, characterized by a high rate of short-duration events, and a slow one, characterized by infrequent events with exponentially distributed residence times. The event counts for both mechanisms are modeled as independent Poisson random variables.

Although the formulation of Ref.~\cite{Sanchez2026} incorporates only one explicit slow retention mechanism, it may still serve as an effective description whenever resolving slow-retention heterogeneity into distinct kinetic classes is unnecessary.
The natural extension is therefore to allow an arbitrary number of independent slow mechanisms, each characterized by its own retention-event rate and release rate, so that such heterogeneity can be represented explicitly within the same stochastic-diffusive framework.

Deriving a formulation of this extension that preserves the practical computability of the original time-domain model poses a significant challenge, because a direct treatment leads to multiple nested summations over the event counts of all mechanisms, and the resulting conditional distributions involve convolutions of Gamma densities with heterogeneous scales, a structure whose computational cost grows rapidly with the number of mechanisms and the truncation level, rendering direct evaluation impractical in the general case.

The present work overcomes this obstacle. An analytic expression for the multi-mechanism case is derived, together with a fast evaluation scheme. Analytical derivatives of the peak-shape function with respect to all model parameters are also obtained, and each can be evaluated at a cost comparable to that of the function itself, which makes efficient gradient-based fitting feasible without sacrificing mechanistic detail.

\section{Theory}
Following Ref.~\cite{Sanchez2026}, the total residence time \(T\) of an analyte particle in the column is decomposed as
\begin{equation}
	T = T_G + T_s,
\end{equation}
where \(T_G\sim\mathcal{N}(\mu_G,\sigma_G^2)\) is a random variable representing the combined contribution of mobile-phase transport and the retention in the stationary phase due to the fast mechanism, and \(T_s\) is an independent random variable representing the additional residence time arising from slow retention events. In the single-mechanism setting, if \(N\sim\mathrm{Poisson}(\Lambda)\) denotes the number of slow retention events, the conditional law of \(T_s\) is
\begin{equation}
	T_s\mid N=n \sim \mathrm{Gamma}(n,\theta), \qquad \mathrm{Gamma}(0,\theta)\equiv\delta,
\end{equation}
with \(\delta\) the Dirac delta. The residence-time PDF takes the form
\begin{equation}\label{eq:base_form}
	f_T(t) = e^{-\Lambda}\sum_{\ell=0}^{\infty}\Omega_\ell\, h_\ell(t),
\end{equation}
with \(h_\ell(t)=\bigl(f_G*f_\Gamma(\cdot;\ell,\theta)\bigr)(t)\), where \(f_G\) is the PDF of \(T_G\), \(f_\Gamma(\cdot;\ell,\theta)\) the Gamma PDF with shape \(\ell\) and scale \(\theta\), and \(\Omega_\ell\) the corresponding expansion coefficients.

The present extension generalizes the model to $M$ independent slow mechanisms.
The retention time due to slow mechanisms is accordingly defined as the sum of independent contributions from each mechanism $j\in\{1,\dots,M\}$:
\begin{equation}
	T_s = \sum_{j=1}^M T_{s,j}, \quad \text{where} \quad T_{s,j} = \sum_{i=1}^{N_j} \tau_{j,i},
\end{equation}
with \(N_j\sim\mathrm{Poisson}(\Lambda_j)\) the number of retention events attributed to mechanism \(j\), and \(\tau_{j,i}\sim\mathrm{Exp}(\beta_j)\) i.i.d.\ random variables representing individual event residence times. Introducing \(\theta_j:=\beta_j^{-1}\), the conditional distribution of \(T_{s,j}\) is
\begin{equation}
	(T_{s,j} \mid N_j = n_j) \overset{d}{=} T_{s,j}^{(n_j)} \sim \mathrm{Gamma}(n_j,\theta_j)
\end{equation}
where $\mathrm{Gamma}(n,\theta)$ denotes a Gamma distribution with integer shape $n$ and scale $\theta$. Furthermore, $T_{s,j}^{(0)} := 0$. That is, mechanism $j$ does not contribute if it yields no retention events ($n_j=0$). Given the event-count vector \(\mathbf{n}:=(n_1,\dots,n_M)\in\mathbb{N}_0^M\), the total conditional retention time due to slow mechanisms is
\begin{equation}
	T_s^{(\mathbf{n})} := \sum_{j=1}^M T_{s,j}^{(n_j)}.
\end{equation}
The extension to multiple mechanisms is formally straightforward, but marginalizing over the independent event counts yields a multi-index Poisson-weighted mixture of Gamma convolutions with heterogeneous scales, which does not reduce to the single-index representation of the one-mechanism case without further manipulation.
The derivation therefore begins by recasting this mixture into a common scale, from which a single-index series representation is recovered.

To this end, \(T_s^{(\mathbf{n})}\) for \(\mathbf{n}\neq\mathbf{0}\) is expressed in terms of Gamma PDFs sharing the common scale \(\theta_1:=\min_j\theta_j\), following the method of Moschopoulos~\cite{Moschopoulos1985}:
\begin{equation}
	f_{T_s^{(\mathbf{n})}}(t) = C(\mathbf{n})\sum_{k=0}^{\infty}\delta_k(\mathbf{n})\,
	f_\Gamma(t;\,\rho(\mathbf{n})+k,\,\theta_1), \quad t>0,
\end{equation}
and zero elsewhere, where
\begin{align}
	\rho(\mathbf{n}) &:= \sum_{j=1}^M n_j \ge 1, \\
	C(\mathbf{n}) &:= \prod_{j=1}^M \left(\frac{\theta_1}{\theta_j}\right)^{n_j},
\end{align}
and \(\delta_k(\mathbf{n})\) satisfies the recurrence
\begin{align}
	\delta_0(\mathbf{n}) &= 1, \\
	\delta_k(\mathbf{n}) &= \frac{1}{k}\sum_{i=1}^k i\,\gamma_i(\mathbf{n})\,\delta_{k-i}(\mathbf{n}),
	\quad k\ge 1,
\end{align}
with
\begin{equation}
	\gamma_i(\mathbf{n}) = \frac{1}{i}\sum_{j=1}^M n_j\left(1-\frac{\theta_1}{\theta_j}\right)^i.
\end{equation}

The marginal PDF \(f_{T_s}\) is obtained by summing over the full state space \(\mathbb{N}_0^M\), weighted by the joint Poisson probability
\begin{equation}
	P(\mathbf{n}) = e^{-\Lambda}\prod_{j=1}^M\frac{\Lambda_j^{n_j}}{n_j!}, \qquad
	\Lambda := \sum_{j=1}^M\Lambda_j.
\end{equation}
Separating the zero-event case \(\mathbf{n}=\mathbf{0}\), for which \(T_s=0\) and hence \(f_{T_s^{(\mathbf{0})}}(t)=\delta(t)\), yields
\begin{align}\label{eq:fTs}
	f_{T_s}(t) &= e^{-\Lambda} \delta(t) \\
	&+ e^{-\Lambda} \sum_{\mathbf{n} \neq \mathbf{0}} \left( \prod_{j=1}^M \frac{\Lambda_j^{n_j}}{n_j!} \right) C(\mathbf{n}) \sum_{k=0}^{\infty} \delta_k(\mathbf{n}) \, f_{\Gamma}(t; \rho(\mathbf{n}) + k, \theta_1)
\end{align}

Since \(T_G\) and \(T_s\) are independent, the residence-time PDF is obtained through the convolution $f_T(t) = (f_G * f_{T_s})(t)$. Following the notation in Ref. \cite{Sanchez2026}, define
\begin{equation}
	h_\ell(t) := \bigl(f_G*f_\Gamma(\cdot;\,\ell,\,\theta_1)\bigr)(t),\quad \ell\ge 1,
\end{equation}
For $\ell=0$, the corresponding contribution of $f_{T_s}$ reduces to a Dirac delta. Thus, by defining $h_0(t):= f_G(t)$ and using the linear property of the convolution operator, it follows that
\begin{equation}
	f_T(t) = e^{-\Lambda}h_0(t)
	+ e^{-\Lambda}\sum_{\mathbf{n}\neq\mathbf{0}}
	\left(\prod_{j=1}^M\frac{\Lambda_j^{n_j}}{n_j!}\right)
	C(\mathbf{n})\sum_{k=0}^{\infty}\delta_k(\mathbf{n})\,h_{\rho(\mathbf{n})+k}(t).
\end{equation}
This expression admits the compact single-index representation
\begin{equation}
	f_T(t) = e^{-\Lambda}\sum_{\ell=0}^{\infty}\Omega_\ell\,h_\ell(t),
\end{equation}
obtained by collecting all contributions to a common index \(\ell=\rho(\mathbf{n})+k\). Since $\mathbf{n} \in \mathbb{N}_0^M \setminus \{\mathbf{0}\}$ implies $\rho(\mathbf{n}) \ge 1$, the index \(\ell\) is strictly positive for all \(k\ge 0\). Thus, the \(\ell=0\) term arises exclusively from the \(\mathbf{n}=\mathbf{0}\) component, giving \(\Omega_0=1\). 

For a given integer $\ell \ge 1$, all contributions from pairs $(\mathbf{n}, k)$ satisfying $\rho(\mathbf{n}) + k = \ell$ are collected. Because $\rho(\mathbf{n}) \ge 1$, the index $k$ is bounded by $k \le \ell - 1$. 
\begin{equation}\label{eq:OmegaL}
	\Omega_\ell = \sum_{\mathbf{n} \in \mathbb{N}_0^M \setminus \{\mathbf{0}\}} \sum_{\substack{k=0 \\ \rho(\mathbf{n})+k=\ell}}^{\ell-1} \left( \prod_{j=1}^M \frac{\Lambda_j^{n_j}}{n_j!} \right) C(\mathbf{n}) \delta_k(\mathbf{n})
	\qquad \ell\ge 1.
\end{equation}

\section{Numerical evaluation}
In the parameter regimes relevant to chromatography, the derived expression for $f_T$ can be evaluated at low computational cost.
 Moreover, the number of slow mechanisms $M$ required in practice is expected to remain small in many cases, given that the single-mechanism model of Ref.~\cite{Sanchez2026} already provides a close approximation to experimental peak shapes.
Two potential computational obstacles nevertheless arise. First, for large $M$ and $\ell$, direct evaluation of the combinatorial coefficients $\Omega_\ell$ via Eq.~\eqref{eq:OmegaL} becomes prohibitively expensive. 
Second, although Ref.~\cite{Sanchez2026} provided an exact and computationally efficient representation of $h_\ell(t)$ in terms of confluent hypergeometric functions for the single-mechanism case, the present multi-mechanism extension increases the dimensionality of the fitting problem and can require a much larger number of function evaluations during optimization. 
Under these conditions, a faster evaluation strategy becomes important. The analysis below therefore derives representations for both $\Omega_\ell$ and $h_\ell$ that are rapidly evaluable, together with analytical derivatives for efficient gradient-based fitting.

\subsection{Evaluation of $\Omega_\ell$}
From Eqs. \eqref{eq:fTs} and \eqref{eq:OmegaL} it follows that 
\begin{equation}
f_{T_s}(t) = e^{-\Lambda}\delta(t) + e^{-\Lambda}\sum_{\ell\ge1}\Omega_\ell\,f_\Gamma(t;\ell,\theta_1)
\end{equation}
Its Laplace transform is
\begin{equation}
\mathcal{L}_{T_s}(s) = e^{-\Lambda}\left(1+\sum_{\ell\ge1}\Omega_\ell(1+\theta_1 s)^{-\ell}\right)
\end{equation}
This follows from the linearity of the Laplace transform, $\mathcal{L}\{\delta\}=1$ and $\mathcal{L}\{f_\Gamma(\cdot;\ell,\theta_1)\}=(1+\theta_1 s)^{-\ell}$.

As $T_s$ represents the sum of $M$ independent compound Poisson components with exponential residence times, its Laplace transform is also
\begin{equation}\label{eq:TsLaplaceKnown}
\mathcal{L}_{T_s}(s) = \exp\!\left(-\sum_{j=1}^M \Lambda_j\,\frac{\theta_j s}{1+\theta_j s}\right), \qquad \Re(s)>0.
\end{equation}
Let $u := (1+\theta_1 s)^{-1}$, then
\begin{equation}\label{eq:Fudeff}
	F(u) := e^{\Lambda}\mathcal L_{T_s}(s(u)) = 1+\sum_{\ell\ge1}\Omega_\ell u^\ell,
\end{equation}
so $\Omega_\ell$ are the Maclaurin coefficients of $F$. Without loss of generality, let $\theta_1=\min_j\theta_j$ and define
\begin{equation}
a_j := 1-\frac{\theta_1}{\theta_j}\in[0,1).
\end{equation}
Substituting $s=(1-u)/(\theta_1 u)$ into the quotient in Eq. \eqref{eq:TsLaplaceKnown} and using the definition of the $a_j$ yields
\begin{equation*}
\frac{\theta_j s}{1+\theta_j s} = \frac{1-u}{1-a_j u},
\end{equation*}
and therefore
\begin{equation}
F(u) = \exp\!\left(\sum_{j=1}^M \Lambda_j\left[1-\frac{1-u}{1-a_j u}\right]\right).
\end{equation}

The bracket simplifies to $(1-a_j)u/(1-a_j u)$. For $\Re(s)>0$, $0<|u|<1$, and since $0\le a_j<1$, it follows that $|a_j u|<1$. Thus, the expression in the bracket can be expanded in geometric series, which after reindexing yields
\begin{equation}
	F(u) = \exp\!\left(\sum_{j=1}^M \Lambda_j (1-a_j)\sum_{m\ge1}a_j^{m-1}u^m\right)
\end{equation}
This is equivalently written as
\begin{align}
F(u) &= \exp\!\left(\sum_{m\ge1} b_m u^m\right)\\	
b_m &:= \sum_{j=1}^M \Lambda_j(1-a_j)a_j^{m-1},\;\; m\ge1 
\end{align}
Differentiating the latter expression for $F(u)$ yields
\begin{equation}
F'(u) = F(u)\left(\sum_{m\ge1} m b_m u^{m-1}\right).
\end{equation}
Replacing $F(u)$ and $F'(u)$ according to Eq. \eqref{eq:Fudeff} with $\Omega_0=1$
\begin{equation}
	\sum_{\ell\ge1} \ell \Omega_\ell u^{\ell-1} = \left(\sum_{l\ge0}\Omega_l u^l\right) \left(\sum_{m\ge1} m b_m u^{m-1}\right).
\end{equation}
The scalar coefficients of every power of $u$ on both sides of the equation must be identical.
 On the right-hand side, the product of terms containing $u^l$ and $u^{m-1}$ yields $u^{l+m-1}$. 
Collecting all terms that produce the power $u^{\ell-1}$ requires the indices to satisfy the constraint $l + m - 1 = \ell - 1$, which simplifies to $l = \ell - m$. Because $l \ge 0$ and the summation over $m$ begins at $m = 1$, $m$ is restricted to $1 \le m \le \ell$ to construct the $u^{\ell-1}$ term.

Equating the isolated coefficients from both sides establishes the recurrence relation:
\begin{equation}
\Omega_\ell=\frac{1}{\ell}\sum_{m=1}^{\ell} m\,b_m\,\Omega_{\ell-m}, \qquad \ell\ge1.
\end{equation}

\subsection{Evaluation of $h_\ell$}

Let $f_G(t)=\mathcal N(t;\mu_G,\sigma_G^2)$, and denote by $f_\Gamma(t;\ell,\theta_1)$ the Gamma density in the scale parametrization, with integer shape $\ell\ge1$ and scale $\theta_1$ (rate $\theta_1^{-1}$). For $\ell\ge1$, the convolution
\[
h_\ell(t)=\bigl(f_G * f_\Gamma(\cdot;\ell,\theta_1)\bigr)(t)
\]
admits the integral representation
\[
h_\ell(t)=\int_0^\infty \frac{1}{\sigma_G\sqrt{2\pi}}
\exp\!\left[-\frac{(t-s-\mu_G)^2}{2\sigma_G^2}-\frac{s}{\theta_1}\right]\,
\frac{s^{\ell-1}}{\Gamma(\ell)\,\theta_1^{\ell}}\,ds .
\]
Expanding the argument of the exponential function yields
\[
-\frac{(t-s-\mu_G)^2}{2\sigma_G^2}-\frac{s}{\theta_1}
=
-\frac{(t-\mu_G)^2}{2\sigma_G^2}
-\frac{s^2}{2\sigma_G^2}
-s\!\left(\frac{1}{\theta_1}-\frac{t-\mu_G}{\sigma_G^2}\right).
\]
This quadratic dependence motivates changing the integration variable to $x:=s/(\sigma_G\sqrt2)$, and defining
\begin{equation}
a=\frac{\mu_G+\sigma_G^2/\theta_1-t}{\sigma_G\sqrt2},
\end{equation}
so that the function $h_{\ell\ge1}$ satisfies
\begin{equation}\label{eq:hl_Jn_rep}
	h_\ell(t)=
	\exp\!\left[-\frac{(t-\mu_G)^2}{2\sigma_G^2}\right]\,
	\frac{\theta_1^{-\ell}(\sigma_G\sqrt2)^{\ell-1}}{\Gamma(\ell)\sqrt\pi}\;
	J_{\ell-1}(a)
\end{equation}
where
\begin{equation}
J_n(a):=\int_0^\infty x^n e^{-x^2-2ax}\,dx,\qquad n\ge0.
\end{equation}
Efficient evaluation of $J_n$ can be performed via recurrence relations. The values for $n=0$ and $n=1$ are 
\begin{equation}\label{eq:Jn_bases}
	J_0(a)=\frac{\sqrt\pi}{2}\,\mathrm{erfcx}(a),\qquad
	J_1(a)=\frac{1}{2}-a\,J_0(a).
\end{equation}
For $n\geq2$, $J_n(a)$ is rewritten as
\begin{align}\nonumber
J_n(a) &= \int_0^\infty x^{n-1}(x+a) e^{-x^2-2ax}dx - a \int_0^\infty x^{n-1} e^{-x^2-2ax}dx
\end{align}
The second term equals $a J_{n-1}(a)$. 
If integration by parts is applied with
 $w = x^{n-1}$ and $dv = (x+a)\exp\left(-x^2-2ax\right)dx$, the antiderivative $v = -\exp\left(-x^2-2ax\right)/2$ is obtained. Because the boundary term $\left[w v\right]_0^\infty$ vanishes for $n \ge 2$, the remaining term equals $(n-1)J_{n-2}(a)/2$. Therefore,
\begin{equation}
J_n(a) = \frac{n-1}{2} J_{n-2}(a) - a J_{n-1}(a),\qquad n\ge2.
\end{equation}
With $n=\ell-1$, substitution of Eq.~\eqref{eq:hl_Jn_rep} into the recurrence for $J_n$ yields the corresponding recurrence for $h_\ell$:
\begin{equation}
h_\ell(t) = \frac{1}{\ell-1} \left[ \left(\frac{\sigma_G}{\theta_1}\right)^2 h_{\ell-2}(t) + \left( \frac{t - \mu_G - \sigma_G^2/\theta_1}{\theta_1} \right) h_{\ell-1}(t) \right]
\end{equation}
Although the integral representation in Eq. \eqref{eq:hl_Jn_rep} is strictly defined for $h_{\ell \ge 1}$, this linear recurrence remains valid when extended down to $h_0$. To verify this, express $h_2(t)$ using Eqs. \eqref{eq:hl_Jn_rep} and \eqref{eq:Jn_bases}:
\begin{equation}
h_2(t) = \exp\!\left[-\frac{(t-\mu_G)^2}{2\sigma_G^2}\right] \frac{\theta_1^{-2}\sigma_G\sqrt2}{\sqrt\pi} \left[ \frac{1}{2} - a J_0(a) \right]
\end{equation}

Distributing the terms, recognizing the definition of $h_1(t)$ from Eq.~\eqref{eq:hl_Jn_rep}, and substituting $a$ with its definition yields
\begin{equation}
h_2(t) = \left(\frac{\sigma_G}{\theta_1}\right)^2 f_G(t) + \left( \frac{t - \mu_G - \sigma_G^2/\theta_1}{\theta_1} \right) h_1(t)
\end{equation}
This identically matches the general recurrence relation for $\ell=2$ if $h_0(t) = f_G(t)$.
Consequently, ${h_{\ell}(t)}$ can be generated for $\ell>1$ from $h_0(t)=f_G(t)$ and $h_1(t)$ using the recurrence above. 

\subsubsection{Implementation note}
For $a\ll0$, $\mathrm{erfcx}(a)$ may overflow. In that case, because $\mathrm{erfcx}(a)=\exp(a^2)\mathrm{erfc}(a)$ the unscaled version should be used and the remaining exponential moved to the prefactor, yielding
\begin{equation}
h_1(t)=\frac{1}{2\theta_1} \exp\!\left[-\frac{t-\mu_G}{\theta_1} + \frac{1}{2}\left(\frac{\sigma_G}{\theta_1}\right)^2\right] \mathrm{erfc}(a)
\end{equation}

\subsection{Evaluation of partial derivatives of $\Omega_\ell$}

To implement gradient-based optimization algorithms, the analytical derivatives of the scalar weights $\Omega_\ell$ with respect to the kinetic parameters $\Lambda_j$ and $\theta_j$ are required. Although $\Omega_\ell$ was introduced above as a coefficient of the series expansion, it can be regarded as a smooth function of the parameters $\{\Lambda_j,\theta_j\}$. Differentiating the recurrence relation for $\Omega_\ell$ with respect to an arbitrary parameter $p \in \{\Lambda_j, \theta_j\}$ yields a secondary linear recurrence:
\begin{equation}
\frac{\partial \Omega_0}{\partial p} = 0, \quad \frac{\partial \Omega_\ell}{\partial p} = \frac{1}{\ell} \sum_{m=1}^{\ell} m \left[ \frac{\partial b_m}{\partial p} \Omega_{\ell-m} + b_m \frac{\partial \Omega_{\ell-m}}{\partial p} \right], \quad \ell \ge 1
\end{equation}

Because the derivative $\partial \Omega_{\ell-m} / \partial p$ is already known from the preceding recursive steps, the computational effort is strictly confined to evaluating the derivatives of the generating sequence $b_m$. Since $b_m$ is composed of elementary algebraic operations, its analytical gradients are obtained directly via the chain rule.

The explicit expressions required to compute the Jacobian are summarized below. The derivatives with respect to the Poisson parameters are:
\begin{equation}
\frac{\partial b_m}{\partial \Lambda_j} = (1 - a_j) a_j^{m-1}, \qquad 1 \le j \le M
\end{equation}

For the individual scale parameters $\theta_j$ ($j \ge 2$), the derivative is:
\begin{equation}
\frac{\partial b_m}{\partial \theta_j} = \Lambda_j \frac{\theta_1}{\theta_j^2} \left[ (m-1)a_j^{m-2} - m a_j^{m-1} \right], \qquad 2 \le j \le M	
\end{equation}
Finally, the global base scale $\theta_1$ must also be optimized. The term corresponding to $j=1$ in the summation of $b_m$ evaluates to $\Lambda_1$ for $m=1$ and strictly $0$ for $m \ge 2$; it is therefore independent of $\theta_1$. Consequently, its gradient reduces to a summation over the slower mechanisms ($j \ge 2$):
\begin{equation}
	\frac{\partial b_m}{\partial \theta_1} = \sum_{j=2}^M \frac{\Lambda_j}{\theta_j} \left[ m a_j^{m-1} - (m-1)a_j^{m-2} \right]
\end{equation}

During parameter estimation, these expressions allow the gradients of $\Omega_\ell$ to be evaluated concurrently with the forward recurrence, circumventing the numerical instabilities of finite-difference approximations.

\subsection{Evaluation of derivatives $h_\ell$}

The analytical derivatives of the $h_\ell$ with respect to the parameters $\mu_G$, $\sigma_G$, and $\theta_1$ can be obtained efficiently through differentiation under the integral sign.
Since $\mu_G$ and $\sigma_G$ are parameters only of $f_G$, whereas $\theta_1$ is a parameter only of $f_\Gamma$, the derivative of $h_\ell=f_G*f_\Gamma$ with respect to any of these parameters is obtained by differentiating only the factor that depends on that parameter.
This avoids explicit differentiation of the recurrence relations for $h_\ell$.

Since $f_G$ depends on $t$ and $\mu_G$ only through $t-\mu_G$: $\partial f_G / \partial \mu_G = -\partial f_G / \partial t$. The temporal derivative of the Gamma density in the scale parametrization is $\partial f_\Gamma(t;\ell,\theta_1) / \partial t = \theta_1^{-1} [f_\Gamma(t;\ell-1,\theta_1) - f_\Gamma(t;\ell,\theta_1)]$ for $\ell \ge 1$. Applying the convolution operator yields:
\begin{equation}
\frac{\partial h_\ell(t)}{\partial \mu_G} = \frac{1}{\theta_1} \left[ h_\ell(t) - h_{\ell-1}(t) \right], \qquad \ell \ge 1	
\end{equation}
For $\ell=0$, direct differentiation of the base Gaussian yields:
\begin{equation}
\frac{\partial h_0(t)}{\partial \mu_G} = \frac{t-\mu_G}{\sigma_G^2} h_0(t)	
\end{equation}

For the dispersion parameter $\sigma_G$, the Gaussian density satisfies $\partial f_G / \partial \sigma_G = \sigma_G \partial^2 f_G / \partial t^2$. Taking the second temporal derivative of the Gamma density and translating it through the convolution integral:
\begin{equation}
\frac{\partial h_\ell(t)}{\partial \sigma_G} = \frac{\sigma_G}{\theta_1^2} \left[ h_{\ell-2}(t) - 2h_{\ell-1}(t) + h_\ell(t) \right], \qquad \ell \ge 2
\end{equation}
For $\ell = 0$ and $\ell = 1$, the derivatives are evaluated directly:
\begin{align}
\frac{\partial h_0(t)}{\partial \sigma_G} &= \left( \frac{(t-\mu_G)^2}{\sigma_G^3} - \frac{1}{\sigma_G} \right) h_0(t)\\
\frac{\partial h_1(t)}{\partial \sigma_G} &= \frac{\sigma_G}{\theta_1^2} \left[ h_1(t) - h_0(t) \right] - \frac{t-\mu_G}{\theta_1\sigma_G} h_0(t)
\end{align}

For the scale parameter $\theta_1$, the derivative of the Gamma density satisfies $\partial f_\Gamma / \partial \theta_1 = \theta_1^{-1} \ell [f_\Gamma(t;\ell+1,\theta_1) - f_\Gamma(t;\ell,\theta_1)]$. Convolving this identity with $f_G$ yields:
\begin{equation}
\frac{\partial h_\ell(t)}{\partial \theta_1} = \frac{\ell}{\theta_1} \left[ h_{\ell+1}(t) - h_\ell(t) \right], \qquad \ell \ge 1	
\end{equation}
For $\ell=0$, the term $h_0=f_G$ does not depend on $\theta_1$, so that $\partial h_0/\partial \theta_1 = 0$.

\subsection{Evaluation of partial derivatives of the PDF}

To compute the Jacobian of the probability density function with respect to the system parameters, the total derivative is assembled piecewise. Let $f_T^{(L)}(t)$ denote the series approximation truncated at order $L$. The partial derivatives are separated into explicit cases for clarity and computational efficiency.

For the Poisson expected event counts $\Lambda_j$ ($1 \le j \le M$):
\begin{equation}
\frac{\partial f_T^{(L)}(t)}{\partial \Lambda_j} = e^{-\Lambda} \sum_{\ell=0}^{L} \left( \frac{\partial \Omega_\ell}{\partial \Lambda_j} - \Omega_\ell \right) h_\ell(t)	
\end{equation}

For the individual scale parameters $\theta_j$ ($j \ge 2$):
\begin{equation}
\frac{\partial f_T^{(L)}(t)}{\partial \theta_j} = e^{-\Lambda} \sum_{\ell=0}^{L} \frac{\partial \Omega_\ell}{\partial \theta_j} h_\ell(t)	
\end{equation}

For the global base scale parameter $\theta_1$:
\begin{align}
\frac{\partial f_T^{(L)}(t)}{\partial \theta_1} &= e^{-\Lambda} \sum_{\ell=0}^{L} \frac{\partial \Omega_\ell}{\partial \theta_1} h_\ell(t)\\
&+ e^{-\Lambda} \sum_{\ell=1}^{L} \Omega_\ell \frac{\ell}{\theta_1} \left[ h_{\ell+1}(t) - h_\ell(t) \right]
\end{align}
Evaluating the gradient with respect to $\theta_1$ at truncation level $L$ requires computing the $h_\ell$ up to order $L+1$. Furthermore, because the parameter space is restricted by the definition $\theta_1 := \min_j \theta_j$, the objective function is not smoothly differentiable at boundaries where two or more scale parameters coincide. 
Consequently, the fitting implementation used here relies on a reparametrization that enforces the strict ordering $\theta_1 < \theta_2 < \cdots < \theta_M$, thereby excluding cases in which two or more scale parameters coincide from the optimization domain.

The expansion weights $\Omega_\ell$ do not depend on $\mu_G$ or $\sigma_G$, so the chain rule reduces to:
\begin{align}
\frac{\partial f_T^{(L)}(t)}{\partial \mu_G} &= e^{-\Lambda} \sum_{\ell=0}^{L} \Omega_\ell \frac{\partial h_\ell(t)}{\partial \mu_G}\\
\frac{\partial f_T^{(L)}(t)}{\partial \sigma_G} &= e^{-\Lambda} \sum_{\ell=0}^{L} \Omega_\ell \frac{\partial h_\ell(t)}{\partial \sigma_G}
\end{align}

\section{Practical scope}
This section addresses two issues relevant to the practical use of the formulation.
First, the expression is fitted to three literature peaks in order to assess its practical performance, using the exponentially modified Gaussian as a widely used reference model\cite{Kalambet2011,naish1988exponentially}, and to examine how fit quality changes with the number of slow mechanisms $M$.
Second, the computational cost of evaluating the PDF and its analytical Jacobian is compared with that of alternative evaluation strategies.

\subsection{Illustrative fits to literature peaks}
The derived expression was fitted to three peaks taken from the literature.
These are not intended to be representative of chromatographic peaks in general. They were selected from the available literature according to four operational criteria: accessible raw time-signal data, acceptable signal-to-noise ratio, absence of evident peak overlap, and sufficiently pronounced tailing for a useful comparison.
The data for Case~I were obtained from the dataset associated with Refs.~\citenum{paper-caseI} and \citenum{data-caseI}, whereas those for Cases~II and III were obtained from the dataset associated with Ref.~\citenum{paper-caseII} and \citenum{data-caseII}.

All peak intensities were normalized to a maximum height of 100, so that residuals and RMSE values can be read directly as percentages of the peak height.
For each case and value of $M$ considered, the minimum truncation level required at the final fitted parameters was determined so that the fractional area lost by truncation does not exceed $10^{-4}$. The corresponding values of this minimum required $L$ and of the RMSE are reported in Table~\ref{tab:litfits}.

\vspace{0.5cm}\noindent
\begin{minipage}{\columnwidth}
	\captionsetup{type=table}
	\centering
	\scriptsize
	\setlength{\tabcolsep}{5pt}
	\renewcommand{\arraystretch}{1.08}
	\begin{tabular*}{0.96\columnwidth}{@{\extracolsep{\fill}}l cccc@{}}
		\toprule
		& EMG & $M=1$ & $M=2$ & $M=3$ \\
		\midrule
		\multicolumn{5}{@{}l}{\scshape{Case~I}} \\[2pt]
		RMSE & 5.57 & 0.29 & 0.14 & 0.14 \\
		$L_{\min}$ & -- & 5 & 159 & 383 \\
		\midrule
		\multicolumn{5}{@{}l}{\scshape{Case~II}} \\[2pt]
		RMSE & 0.51 & 0.31 & 0.08 & 0.05 \\
		$L_{\min}$ & -- & 5 & 32 & 171 \\
		\midrule
		\multicolumn{5}{@{}l}{\scshape{Case~III}} \\[2pt]
		RMSE & 0.43 & 0.31 & 0.03 & 0.03 \\
		$L_{\min}$ & -- & 6 & 29 & 29 \\
		\bottomrule
	\end{tabular*}
	\captionof{table}{RMSE (in percent-of-height units) and minimum truncation level \(L_{\min}\) required at the final fitted parameters to keep the fractional area lost by truncation below \(10^{-4}\). In all cases, peak intensities were normalized to a maximum height of 100.\vspace{0.5cm}}
	\label{tab:litfits}
\end{minipage}

Results are reported up to $M=3$, as the RMSE ceased to improve significantly at or before this value in all three cases. The cutoff was independently confirmed by the Akaike information criterion.
Figure~\ref{fig:litfits} represents, for each case, the fitted profiles and the corresponding residuals. The residual plots are included because systematic deviations between the fitted expression and the data may not be discernible from the fitted curves alone. For each case, the curves shown are restricted to values of $M$ for which there is a practically significant improvement relative to the preceding one. For visual clarity, the peak-profile panels display regularly subsampled experimental data points, whereas the fits and the residual plots are based on the full data.

In Case~I (Fig.~\ref{fig:litfits}), the EMG leaves large structured residuals over a substantial part of the profile and does not reproduce the central region and the extended tail simultaneously (RMSE\,$=$\,5.57). 
With the present model, the RMSE decreases to 0.29 at $M=1$ and to 0.14 at $M=2$.
Under the present fitting protocol, increasing to $M=3$ does not reduce the RMSE further.
In this case, describing the full peak profile requires more flexibility than that provided by the EMG, but not necessarily a large value of $M$.

In Case~II (Fig.~\ref{fig:litfits}), the dependence of the fitting error on $M$ follows a different pattern.
At $M=1$ the RMSE is already below that of the EMG (0.31 vs.\ 0.51), but introducing a second slow mechanism produces a further reduction by roughly a factor of four (RMSE\,$=$\,0.08 at $M=2$).
At $M=3$, the RMSE decreases further to 0.05.

In Case~III (Fig.~\ref{fig:litfits}), the $M=2$ fit reduces the RMSE by an order of magnitude relative to $M=1$ (0.03 vs.\ 0.31). 
At $M=3$, the RMSE changes only marginally, while one fitted $\Lambda_j$ is driven to a value close to zero. Under the present parametrization, the additional mechanism therefore becomes effectively inactive, so the $M=3$ fit does not provide a practically distinct description from the $M=2$ fit.

For these three cases, every tested instance of the present expression with $M\ge1$ yielded lower RMSE than the EMG, by factors ranging from about 1.4 to 40. This outcome cannot be assumed to hold universally since the present model is not a generalization of the EMG.
Absence of improvement at higher~$M$ should be read as an empirical result under the present fitting protocol, not as proof that no better parameter set exists.

\end{multicols}

\begin{figure}[ht]
	\centering
	 \includegraphics[width=0.94\textwidth]{img.pdf}
\caption{Illustrative fits to three literature peaks
	(see text for details on each case).
	Upper panels: experimental data (black crosses, regularly
	subsampled for clarity) and fitted profiles.
	Lower panels: residuals (dots), computed from the full data.
	Fitted profiles and their corresponding residuals:
	EMG (\,\usebox{\legendEMG}),
	$M=1$ (\,\usebox{\legendMone}\,),
	$M=2$ (\,\usebox{\legendMtwo}\,),
	$M=3$ (\,\usebox{\legendMthree}\,).
	Only model orders producing a visually appreciable improvement
	are shown.
	All intensities are normalized to a maximum height of~100.}
	\label{fig:litfits}
\end{figure}

\begin{multicols}{2}
\raggedcolumns
\subsection{Computational performance}

All timing measurements reported below were obtained from a serial
Fortran implementation compiled with \texttt{gfortran} and executed on a
desktop computer equipped with an AMD Ryzen~5 5600 processor.

The computational cost of the present procedure for evaluating the sequence $h_0,\dots,h_L$ was compared directly with that of the route based on confluent hypergeometric functions presented in Ref.~\cite{Sanchez2026}.
For that route, these special functions were evaluated with the GNU Scientific Library (GSL)\cite{GSL}.
For the comparison, parameter values were taken from the fitted models for Case~I with $M=2$, Case~II with $M=3$, and Case~III with $M=2$. Table~\ref{tab:kernel-speed} contains timing ratios for the evaluation of the full sequence $h_0,\dots,h_L$ over time grids of size $n_t$ and truncation level $L$. The present procedure was faster in all cases tested, by approximately two to four orders of magnitude. Similar results were obtained for other representative chromatographic parameter sets. 
The speedup factor increased with the truncation level $L$. In the present scheme, $h_0$ and $h_1$ are computed directly, whereas each additional $h_{\ell}$ for $\ell>1$ is generated by the recurrence relation at a marginal cost that is independent of $\ell$. By contrast, the alternative route involves repeated evaluation of substantially more expensive special functions whose per-call evaluation cost may vary more appreciably with the parameter values. 
The reported ratios should therefore be understood as representative of parameter ranges typically encountered in practice.

\begin{center}
	\captionsetup{type=table}
	\scriptsize
	\setlength{\tabcolsep}{4pt}
	\renewcommand{\arraystretch}{1.08}
	\begin{tabular*}{0.94\columnwidth}{@{\extracolsep{\fill}}ccccc@{}}
		\toprule
		& & \multicolumn{3}{c}{Speedup factor} \\
		\cmidrule(lr){3-5}
		$n_t$ & $L$ & Case I & Case II & Case III \\
		\midrule
		$10^2$ & 10   & $2.35\times10^{2}$ & $3.36\times10^{2}$ & $4.20\times10^{2}$ \\
		$10^2$ & 100  & $2.52\times10^{3}$ & $3.81\times10^{3}$ & $3.78\times10^{3}$ \\
		$10^2$ & 1000 & $1.36\times10^{4}$ & $1.58\times10^{4}$ & $1.60\times10^{4}$ \\
		\midrule
		$10^4$ & 10   & $2.27\times10^{2}$ & $3.04\times10^{2}$ & $3.97\times10^{2}$ \\
		$10^4$ & 100  & $1.92\times10^{3}$ & $2.89\times10^{3}$ & $2.79\times10^{3}$ \\
		$10^4$ & 1000 & $5.01\times10^{3}$ & $5.50\times10^{3}$ & $5.51\times10^{3}$ \\
		\bottomrule
	\end{tabular*}
	\captionof{table}{Ratio of execution times between the previous route requiring confluent hypergeometric functions and the evaluation scheme presented here for evaluating the sequence $h_0,\dots,h_L$ on $n_t$ time points. Here, $L$ denotes the truncation level. Case I, Case II, and Case III correspond, respectively, to the fitted parameter sets from Case~I with $M=2$, Case~II with $M=3$, and Case~III with $M=2$. All benchmarks were obtained under identical compilation conditions.}
	\label{tab:kernel-speed}
\end{center}

For peak fitting, the relevant measure of computational cost is the elapsed time required to evaluate the PDF and its Jacobian jointly on a grid of $n_t$ time points.
Table~\ref{tab:pdf-jac-speed} reports the evaluation times for this task using the analytical Jacobian, along with the speedup relative to a forward finite-difference Jacobian of the same PDF.
In the tested cases, the analytical Jacobian evaluation was faster with speedup factors ranging from approximately 1 to 3.
In addition, evaluating the Jacobian analytically avoids the errors inherent in finite-difference approximations.

For the truncation levels required in practice, the per-point cost of a joint PDF and Jacobian evaluation is of the order of tens of nanoseconds
to a few microseconds, depending on $M$ and $L$. 
These evaluation times are well suited to routine peak fitting.

The calculations reported here were performed using a Python interface to the Fortran backend mentioned above, with fitting routines carried out using SciPy\cite{2020SciPy-NMeth}.
In this implementation, the positivity constraints on $\sigma_G$, $\Lambda_j$, and $\theta_j$, together with the ordering constraint $\theta_1<\dots<\theta_M$, are enforced by a smooth variable transformation based on the softplus function.
The Jacobian of this transformation is also available in analytic form, and it is combined with that of the PDF through the chain rule.
This allows the use of unconstrained optimization algorithms while preserving analytical gradients with respect to the transformed variables.
The code used for these calculations is publicly
available~\cite{STEPcode}.

\noindent
\begin{minipage}{\columnwidth}
	\captionsetup{type=table}
	\centering
	\scriptsize
	\setlength{\tabcolsep}{4pt}
	\renewcommand{\arraystretch}{1.08}
	\begin{tabular*}{0.96\columnwidth}{@{\extracolsep{\fill}}cc cc cc@{}}
		\toprule
		& & \multicolumn{2}{c}{$n_t=10^2$} & \multicolumn{2}{c}{$n_t=10^4$} \\
		\cmidrule(lr){3-4}\cmidrule(lr){5-6}
		$M$ & $L$ & $t_{\mathrm{anal}}$ (s) & Speedup & $t_{\mathrm{anal}}$ (s) & Speedup \\
		\midrule
		1 & 10   & $5.77\times10^{-6}$ & $2.23\times$ & $1.23\times10^{-3}$ & $1.01\times$ \\
		1 & 100  & $3.60\times10^{-5}$ & $1.51\times$ & $3.84\times10^{-3}$ & $1.05\times$ \\
		1 & 1000 & $1.37\times10^{-3}$ & $1.62\times$ & $2.96\times10^{-2}$ & $1.14\times$ \\
		2 & 10   & $6.39\times10^{-6}$ & $2.83\times$ & $1.38\times10^{-3}$ & $1.27\times$ \\
		2 & 100  & $4.90\times10^{-5}$ & $1.55\times$ & $4.62\times10^{-3}$ & $1.24\times$ \\
		2 & 1000 & $2.21\times10^{-3}$ & $1.39\times$ & $3.83\times10^{-2}$ & $1.26\times$ \\
		3 & 10   & $7.16\times10^{-6}$ & $3.10\times$ & $1.47\times10^{-3}$ & $1.45\times$ \\
		3 & 100  & $6.30\times10^{-5}$ & $1.52\times$ & $4.94\times10^{-3}$ & $1.42\times$ \\
		3 & 1000 & $2.97\times10^{-3}$ & $1.31\times$ & $4.31\times10^{-2}$ & $1.40\times$ \\
		\bottomrule
	\end{tabular*}
	\captionof{table}{Evaluation time for the joint evaluation of the PDF and its analytical Jacobian (\(t_{\mathrm{anal}}\)), and speedup relative to the corresponding joint evaluation using a forward finite-difference Jacobian. Synthetic time grids were generated from the fitted physical parameters of Case~II for \(M=1,2,\) and \(3\).}
	\label{tab:pdf-jac-speed}
\end{minipage}

\section{Conclusions}
A time-domain analytic expression has been derived by extending the single-slow-mechanism stochastic-diffusive formulation of Ref.~\cite{Sanchez2026} to an arbitrary number of independent slow retention mechanisms. The resulting multi-index Poisson-weighted mixture has been recast into a single-index series whose terms are products of scalar weights $\Omega_\ell$ and functions $h_\ell$, a structure analogous to that of the previous formulation restricted to one slow mechanism. In this series, $h_\ell$ retains the same functional form as in that restricted case. Efficient evaluation schemes have been derived for both the weights $\Omega_\ell$ and the functions $h_\ell$. For the latter, the present scheme replaces the previous route based on confluent hypergeometric functions, reducing the computational cost by two to four orders of magnitude in the cases tested. Analytical derivatives with respect to all model parameters have also been obtained and can be evaluated at a comparable computational cost, enabling efficient gradient-based fitting without the drawbacks of finite-difference approximations. In fits to three literature peaks, every tested instance of the present formulation yielded a lower RMSE than the EMG. Furthermore, allowing more than one slow mechanism reduced the fitting error relative to the single-mechanism formulation, with improvements ranging from roughly a factor of two to more than one order of magnitude.

\section*{Acknowledgement}
This work was supported by the National Scientific and Technical Research Council of Argentina (CONICET).
\linespread{1.05} 

\printbibliography
\end{multicols}
\end{document}